# Computational Assessment of Silicon Quantum Gate Based on Detuning Mechanism for Quantum Computing

Tong Wu and Jing Guo, *Senior Member, IEEE*

*Abstract*—Silicon-based quantum computing has the potential advantages of low cost, high integration density, and compatibility with CMOS technologies. The detuning mechanism has been used to experimentally achieve silicon two-qubit quantum gates and programmable quantum processors. In this paper, the scaling behaviors and variability issues are explored by numerical device simulations of a model silicon quantum gate based on the detuning mechanism. The device physics of quantum gates modulation, tradeoff between device speed and quantum fidelity, and impact of variability on the implementation of a quantum algorithm are examined. The results indicate the attractive potential to achieve high speed and fidelity silicon quantum gates with a low operation voltage. To scale up, reducing the device variability and mitigating the variability effect are identified to be indispensable for reliable implementing a quantum computing algorithm with the silicon quantum gates based on the detuning mechanism. A scheme to use the control electronics for mitigating the variability of quantum gates is proposed.

## I. INTRODUCTION

Silicon-based quantum computing has the interesting features of high integration density, low cost, and compatibility with CMOS technologies for future quantum computers [1]-[8]. In silicon-based quantum computing, the spin qubits confined in the silicon quantum dots can be modulated by external electric and magnetic fields to fulfill the single-qubit gates and two-qubit controlled gates. These quantum gates can form a complete set for universal quantum computing [9]-[16]. The key challenges for realizing silicon-based quantum computing include creating sufficiently strong entanglement between spin qubits and reducing decoherence due to the environment for fast and reliable quantum gates, as well as reducing device-to-device variability and developing low-temperature control electronics for integration. A programmable, two-qubit silicon quantum processor has been experimentally demonstrated recently [7], in which the quantum entanglement was modulated by the detuning mechanism, and the Deutsch-Jozsa and Grover algorithms have been implemented. The demonstrated controlled phase gate has a switching time of ~300ns (corresponding to a frequency of ~3 MHz) and a fidelity of ~85% for the controlled-phase (CZ) gate [7]. Furthermore, strong spin-photon coupling in silicon has also been demonstrated recently [17]-[19], which opens an attractive route to achieve long-range coupling and scale up quantum networks based on silicon spin qubits.

With the rapid experimental progress made, it is important to understand the device issues related to the performance potential, variability, and scaling characteristic from a device modeling and simulation perspective. Previous theoretical works on the detuning quantum gates, however, have focused on fundamental physical understandings without directly relating device structural and material parameters to the performance [20], [21]. Simplified approximations were also used, which can fail in a quantitative accuracy test [22], [23]. Also needed, however, is device modeling and simulations that can quantitatively and predictively calculate key device performance metrics. The device simulation approach has been recently applied to resonantly driven silicon quantum gates, which is not based on the detuning mechanism [24]. The objective of this work is to develop and use numerical device simulations to examine the device performance potential, design optimization, and device variability issues for silicon two-qubit quantum gates based on the detuning mechanism [7], [25].

Fig. 1(a) and (b) summarizes the operation principles of a two-qubit CZ gate based on the detuning mechanism [7], [25]. Two quantum dots can be electrostatically defined by the applied gate voltages $V_{g1}$ and $V_{g2}$. The detuning interdot potential difference can be controlled by the gate voltage difference $\Delta V_g = V_{g1}\text{-}V_{g2}$. The applied magnetic field along the z-direction lifts the degeneracy of the triplet energy levels to $\{T_1, T_0, T_{-1}\}$, where the subscripts denote the spin along z-direction, $m_z$, as shown in Fig. 1(b). Both the singly occupied singlet state $S_{11}$ and doubly occupied singlet state $S_{20}$ have $m_z=0$, and they are hybridized. Among all levels, only the doubly occupied state, $S_{20}$ is sensitive to the detuning gate voltage $\Delta V_g$. The states $\{T_1, T_0, S_{11}, T_{-1}\}$ form a basis for two-qubit quantum computing, and the detuning gate voltage controls the energy shift of the $S_{11}$ level with regard to the triplet levels, which leads to a controlled phase operation. By controlling the wide and strength of the detuning voltage pulse, a total phase shift of $\pi$ is

T. Wu and J. Guo are with the Department of Electrical Engineering, University of Florida, Gainesville, FL, 32611-6130 USA. (e-mail: guoj@ufl.edu).

implemented in a CZ gate. The above mechanism applies to a position-independent magnetic field. A magnetic field gradient ($B_1 \neq B_2$) can be applied to make each spin qubit individually addressable in single-gate rotational operations. In this case, the magnetic field gradient mixes the $T_0$ and $S_{11}$ states to form renormalized spin states [21].

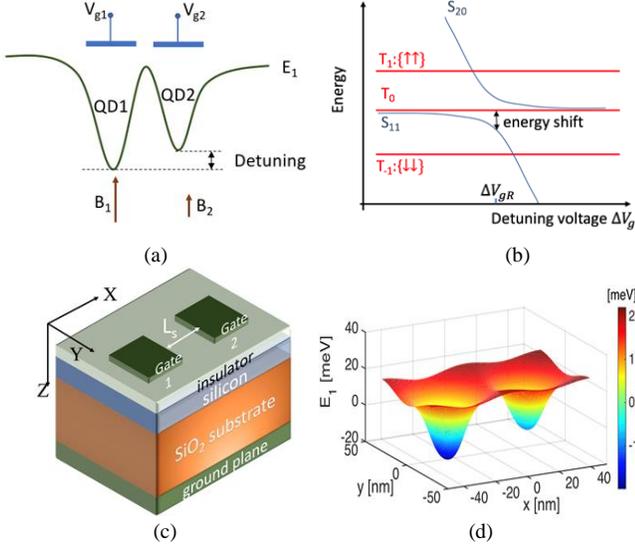

Fig. 1. (a) Schematic band profile (E1) of a CZ gate based on the detuning mechanism. The two quantum dots (QD1 and QD2) are created and detuned by the gate voltages $V_{g1}$ and $V_{g2}$ respectively. $B_1$ and $B_2$ are magnetic fields. (b) Schematic sketch of the energy levels vs. the detuning gate voltage $\Delta V_g = V_{g1} - V_{g2}$ with a position-independent magnetic field. (c) Modeled device structure of the silicon quantum gate on a silicon-on-insulator (SOI) structure. The coordinate system is also shown. Both gate electrodes have a size of 10nm (along the x direction) and 20nm (along the y direction) with a spacing of $L_s$. (d) Simulated subband profile at the average gate voltage of $\bar{V}_g = (V_{g1} + V_{g2})/2 = 60$ mV, $\Delta V_g = 11.806$ mV, and $L_s = 40$ nm.

In section II, a multiscale numerical simulation approach is described to simulate the silicon detuning quantum gates. In section III, the simulation results are presented, and the device physics of detuning, device performance potential, scaling behaviors and variability issues are discussed. Section IV provides the conclusion.

## II. APPROACH

A multiscale simulation approach is used. Numerical device simulations based on the configuration interaction (CI) method are first used to characterize the many-body energy levels and wave functions in the silicon detuning quantum gate device [22]. An effective Hamiltonian is then parameterized, and the Lindblad Master equation [26], [27] is subsequently solved to assess key device performance metrics.

Fig. 1(c) shows the schematic of the modeled device. In the lab experimental demonstrations, multiple gate electrodes were typically used [7], [25]. The modeled device has a more compact structure with two gate electrodes, which can create and detune two entangled quantum dots. Various silicon structures have been used in experimental demonstrations of quantum gates [1], [4], [6], [7], [28]-[30]. A silicon-on-insulator (SOI) structure is simulated here. The silicon film thickness is $t_{si} = 3$ nm, and it is confined along the [100] direction. The SOI film has a 10-nm-thick SiO$_2$ substrate. The top gate insulator has a thickness of $t_{ins} = 3$ nm and a relative dielectric constant of $\kappa = 25$. The silicon film thickness and gate insulator parameters are comparable to the state-of-the-art transistor technologies [31]. A low temperature of $T = 20$ mK is assumed. In the energy range of interest, the valley degeneracy of silicon can be lifted by the interface effects [32], and only one valley is considered. The existence of the higher lying valley states can introduce an undesired valley degree of freedom, and thereby reduce the quantum coherent time. In a silicon MOS device structure, a valley splitting in the range of 300-800 μeV, tunable by the electric field, has been experimentally demonstrated [33], which is larger than the Zeeman splitting in the modeled device. For a device system with a valley splitting smaller than the energy scales of interest, further studies will be needed to quantitatively elucidate the impact of the valley degree of freedom, which is beyond the scope of this paper.

A key task of the numerical simulation is to assess the many-body eigenenergies by using the CI method [34]. The CI method can be progressively more accurate as the size of the basis set increases. Previously, the CI method has previously been used to investigate the GaAs-based double quantum gate (DQD) singlet-triplet qubit, with an approximate quadratic potential [35]. The CI calculations in this paper remove the quadratic potential approximation, which allow a more accurate modeling of the device electrostatics, and they are used to examine the device performance metrics of the silicon-based two-qubit quantum gates. To form a basis set for solving low energy states, we first obtain the quasiparticle wave functions by solving a 3-D Poisson equation and the quasi-particle Schrödinger equation in the absence of electron-electron interaction. For the modeled device structure as shown in Fig. 1(c), a finite-difference mesh was used to discretize the Poisson and Schrödinger equations. We use a mode space approach to solve the 3-D Schrödinger equation, which decouples the 3-D problem to a 1-D Schrödinger equation in the vertical confinement direction and a 2-D Schrödinger equation in the in-plane direction. The mode space approach has been shown to be accurate for a thin SOI film with a uniform thickness [36].

After the wave functions of the lowest quasi-particle states are calculated, the basis set of the CI method is obtained as the Slater-type products of the lowest quasi-particle wave functions [22], [23]. The two-body Hamiltonian can be expressed as,

$$\hat{H}(\vec{r_1}, \vec{r_2}) = \sum_{i=1,2} \hat{h_i}(\vec{r_i}) + \hat{c}(\vec{r_1}, \vec{r_2}) \quad (1)$$

where $\hat{h_i}(\vec{r_i})$ is the quasi-particle Hamiltonian in the absence of electron-electron interaction and $\hat{c}(\vec{r_1}, \vec{r_2})$ is the Coulomb interaction term. In the basis set chosen above, the first term $\hat{h_i}(\vec{r_i})$ is diagonal, and the Coulomb term $\hat{c}(\vec{r_1}, \vec{r_2})$ introduces nondiagonal entries in the Hamiltonian matrix [22]. We use the lowest $N = 8$ quasi-particle wave functions to form a CI basis set of $N^2 = 64$ Slater wave function products. The accuracy of the calculated energy levels is confirmed by progressively further increasing the size of the basis set.

The CI simulations are computationally intensive. For efficient device simulations, a simple effective Hamiltonian

[21], [25] is used to parameterize the CI simulations,

$$H_{eff} = \begin{bmatrix} E_z/2 & 0 & 0 & 0 & 0 \\ 0 & E_{z1}/2 & 0 & 0 & h_c \\ 0 & 0 & -E_{z1}/2 & 0 & h_c \\ 0 & 0 & 0 & -E_z/2 & 0 \\ 0 & h_c & h_c & 0 & U(\Delta V_g) \end{bmatrix} \quad (2)$$

where the basis set is $\{|\uparrow\uparrow\rangle, |\uparrow\downarrow\rangle, |\downarrow\uparrow\rangle, |\downarrow\downarrow\rangle, S_{20}\}$, the Zeeman splittings are $E_z \approx 2\mu_B(B_1 + B_2)$, $E_{z1} \approx 2\mu_B(B_1 - B_2)$, $h_c$ is the tunneling coupling energy, $U(\Delta V_g) = U_0 - \alpha q \Delta V_g$ is the energy of the $S_{20}$ state, $U_0$ is the Hubbard-Coulomb interaction energy, $q$ is the elementary electron charge, and $\alpha$ is a unitless gating efficiency factor. The parameters in the effective Hamiltonian model $h_c$, $U_0$, and $\alpha$ are extracted by fitting to the CI simulations. The resonant condition is defined as $U(\Delta V_{gR}) = 0$, corresponding to $\Delta V_{gR} = \frac{U_0}{\alpha q}$, when the hybridization between the $S_{11}$ and $S_{20}$ states is maximized. As schematically shown in Fig. 1(b), the anticrossing split between the singlet states is $h_c$ at $\Delta V_g = \Delta V_{gR}$. By parameterizing the simulation results to an effective Hamiltonian, the computationally intensive CI simulation only needs to be done once, and the issue of computational cost can be addressed.

Once the effective Hamiltonian $H_{eff}$ is parameterized, the Lindblad Master equation is solved [26], [27], [37]

$$\frac{d\rho(t)}{dt} = \frac{-i}{\hbar}[H_{eff}, \rho(t)] + \sum_{k=1}^{N} \Gamma_k \left( O_k \rho O_k^+ - \frac{1}{2}\{O_k^+ O_k, \rho(t)\} \right) \quad (3)$$

where $\rho(t)$ is the time-dependent density matrix, and the size of the basis set is $N$. The coherent evolution is described by the first term on the right-hand side (RHS) of the equation, and the decoherence is phenomenologically described by the second term on the RHS of the equation, where the $(m, n)$ matrix element of $O_k$ is $O_k(m,n) = \delta(k,m)\delta(k,n)$ with $\delta$ being the Kronecker delta function, $\Gamma_k = \gamma^*$, and $\gamma^*$ is the decoherence rate. By using this form of $O_k$, the phase decoherence mechanism is treated and energy relaxation is neglected [9], as the phase coherence time is practically shorter than the energy relaxation time. The phenomenological decoherence model is used to investigate the effect of the decoherence rate on the quantum fidelity of the gates and implementation of quantum algorithms.

## III. RESULTS AND DISCUSSION

The simulated 2-D subband profile is examined first. Fig. 1(d) shows the subband profile at an average gate voltage of $\overline{V}_g = (V_{g1} + V_{g2})/2 = 60$ mV and a detuning voltage of $\Delta V_g = V_{g1} - V_{g2} \approx 11.806$ mV. The applied gate voltages define the DQDs structure, in which two spin qubits are coupled through a tunneling barrier along the $x$-direction. In contrast to a symmetrically biased quantum gate [37], in a detuning quantum gate, the tunneling barrier height is not modulated. Instead, $\Delta V_g$ is applied to detune the potential between two quantum dots. The size of the quantum dots is mostly controlled by the size of the gate electrodes, and the spacing between the quantum dots are controlled by the spacing of the gate electrodes.

Next, the CI calculation results are presented and parameterized to the effective Hamiltonian. The energy levels from the CI simulations for a DQD spacing of $L_s = 40$ nm, $B_1 = B_2 = 10$ mT are plotted as the symbols in Fig. 2(a). In the presence of a position-independent magnetic field, the degeneracy of the triplet states with $m_z = +1, 0, 1$ is lifted, and the singlet levels are modulated by the detuning gate voltage. Fig. 2(b) and (c) shows the many-body wave functions of the lowest singlet and triplet states, which are symmetric and antisymmetric, respectively. The lowest singlet wave function shows hybridization to $S_{20}$ as the peak at the diagonal line of $x_1 = x_2$. The energy of the $T_0$ state is used as the energy reference $E = 0$. The effective Hamiltonian model describes the CI simulation results well.

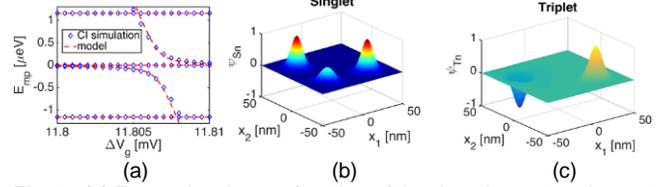

Fig. 2. (a) Energy levels as a function of the detuning gate voltage at a magnetic field of $B_1 = B_2 = 10$ mT from the CI simulations (blue symbols) and the effective Hamiltonian model (red dashed lines). The parameters of the effective Hamiltonian are $\alpha$=0.75, $h_c$=0.34 µeV, and $\Delta V_{gR}$ =11.8066 mV. The wave functions of the lowest (b) singlet state and (c) triplet state are obtained at a detuning gate voltage of $\Delta V_g$ =11.806 mV and an average gate voltage of $\overline{V}_g = 60$ mV. The many-body wave functions are plotted as $\psi(x_1, x_1, y_1 = y_2 = 0, z_1 = z_2 = t_{si}/2)$. $y_1 = y_2 = 0$ and $z_1 = z_2 = t_{si}/2$ define the planes through the centers of the quantum dots, and $t_{si}$ is the silicon film thickness. The modeled device is shown in Fig. 1(c) with a DQD spacing of $L_s = 40$ nm.

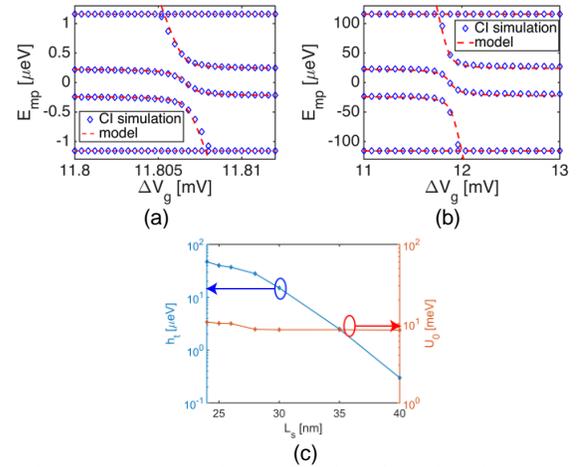

Fig. 3. Lowest energy levels as a function of the detuning gate voltage from the CI simulations (blue symbols) and the effective Hamiltonian model (red dashed lines) for (a) $L_s = 40$ nm, $B_1 = 12$ mT, and $B_2 = 8$ mT with the other parameters same as Fig. 2(a), and for (b) $L_s = 25$ nm, $B_1 = 1.2$ T, and $B_2 = 0.8$ T, with the parameters of the effective Hamiltonian $\alpha = 0.85$, $h_c = 40$ µeV, and $\Delta V_{gR} = 11.88$ mV. (c) Parameterized tunneling coupling $h_c$ and Hubbard energy $U_0$ as a function of the DQD spacing $L_s$.

A magnetic field gradient was often applied in the experiments to make each spin qubit individually addressable [7]. Fig. 3(a) shows the comparison of the energy levels from the CI simulations and the effective Hamiltonian model in the presence of a magnetic field gradient, $B_1 = 12$ mT and $B_2 = 8$ mT. The result shows that the same set of parameters of the effective Hamiltonian still describes the CI simulations well. In the presence of the magnetic field gradient, a basis of $\{|\uparrow\uparrow\rangle, |\widetilde{\uparrow\downarrow}\rangle, |\widetilde{\downarrow\uparrow}\rangle, |\downarrow\downarrow\rangle\}$ states form $\{|00\rangle, |01\rangle, |10\rangle, |11\rangle\}$ states for quantum computing, where the tilde denotes renormalization. Both $|\widetilde{\uparrow\downarrow}\rangle$ and $|\widetilde{\downarrow\uparrow}\rangle$ are hybridized with the $S_{20}$ singlet state and have a phase shift in the quantum gate operation. By applying additional single-qubit gates, the phase shifts can be combined to achieve any controlled $Z_{ij}$ gate by $CZ_{ij}|m,n\rangle = (-1)^{\delta(i,m)\delta(j,n)}|m,n\rangle$ [7].

For a smaller $L_s$=25 nm, the results in Fig. 3(b) indicates that the effective Hamiltonian parameterization still works well. The extracted values for the tunneling coupling $h_c$ and the Hubbard energy $U_0$ are plotted as a function of $L_s$ shown in Fig. 3(c). The tunneling coupling $h_c$ exponentially increases, and $U_0$ remains approximately unchanged as $L_s$ scales down.

The effect of the fluctuation of the gate voltage is studied in the following. It has been shown that charge noise is one of the dominant mechanisms that limit the fidelity of the solid-state quantum gates [37]. The effect can be quantitatively described by the derivative of the exchange between two dots, which is defined as $J$, with regard to the detuning gate voltage $d = |(dJ/d\Delta V_g)/q|$. A larger derivative indicates a larger effect of the gate voltage fluctuation, and an insensitivity voltage $\mathcal{J}$ can be characterized as [37]

$$\mathcal{J} = \frac{|J|}{\sqrt{\sum_i (\partial J/\partial V_{gi})^2}} \quad (4)$$

where the sum index $i$ is over gate electrodes. Fig. 4(a) and (b) shows the derivative parameter $d$ and the insensitive voltage as a function of the detuning gate voltage at $L_s = 40$ nm and 25 nm, respectively. The insensitivity voltage $\mathcal{J}$ decreases as the detuning gate voltage increases closer to $\Delta V_{gR}$, which indicates that operating closer to the resonant condition ($U(\Delta V_g)$=0) is not preferred due to high sensitivity to the gate voltage fluctuation. Fig. 4 indicates that the insensitivity voltage increases by over two orders of magnitude as the DQD spacing $L_s$ reduces from 40 nm to 25 nm due to stronger tunneling coupling.

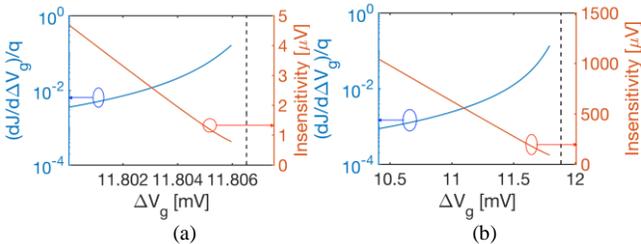

Fig. 4. Sensitivity derivative parameter $d = |(dJ/d\Delta V_g)/q|$ and insensitivity voltage $\mathcal{J}$ versus the detuning gate voltage $\Delta V_g$ simulated for the CZ gate as shown in Fig. 1(c) with (a) $L_s$=40 nm and (b) $L_s$=25 nm. The vertical black dashed lines show the resonant condition ($U(\Delta V_g) = 0$).

After the parameterization, the device performance of the quantum gates can be assessed by solving the Master equation with the effective Hamiltonian. The important quantum gate performance metrics include switching speed, quantum fidelity, and switching gate voltage. We first examine a modeled CZ gate with a silicon DQD spacing of $L_s = 40$ nm, as shown in Fig. 5(a). The gate fidelity and the switching frequency, defined as the inverse of the switching time, are plotted as a function of the gate detuning voltage $\Delta V_g$. When $\Delta V_g$ increases in the range of $\Delta V_g < \Delta V_{gR}$, the device operates closer to the resonant condition, $U(\Delta V_g) = 0$, as shown by the top axis. A larger energy shift results in a faster speed. For a gate bias with $\frac{U(\Delta V_g)}{h_c} = 40$, the switching frequency is ~6 MHz, which is in the same order of magnitude as the speed demonstrated in a recent experiment [7].

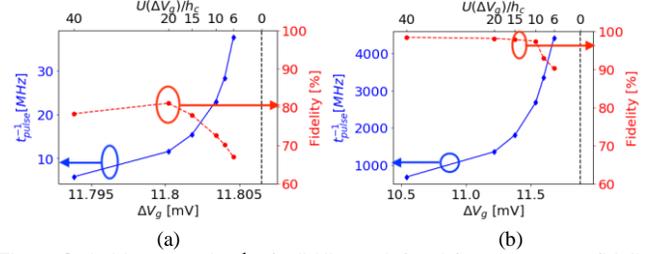

Fig. 5. Switching speed $t_{pulse}^{-1}$ (solid line to left axis) and quantum fidelity (dashed lines to right axis) versus the detuning gate voltage $\Delta V_g$ for a CZ gate as shown in Fig. 1(c) with (a) $L_s = 40$ nm, $B_1 = 12$ mT, and $B_2 = 8$ mT, and (b) $L_s = 25$ nm, $B_1 = 1.2$ T, and $B_2 = 0.8$ T. The top axis shows the corresponding normalized value of $U(\Delta V_g)/h_c$ as in (2). The vertical dashed lines show the resonant condition.

To examine the impact of dephasing on the quantum gate fidelity, a simple, phenomenological model is used to describe the effect of dephasing, in which the total dephasing rate is expressed as

$$\gamma^* = \gamma_0 + \gamma_1(\Delta V_g) \quad (5)$$

where $\gamma_1(\Delta V_g)$ is the rate due to the charge-noise dephasing, and $\gamma_0 = (1\ \mu s)^{-1}$ [7] describes other dephasing mechanisms not sensitive to the detuning gate voltage, such as the hyperfine interaction. Note that longer hyperfine dephasing times can be achieved by using isotope purified $^{28}$Si. The charge-noise dephasing rate $\gamma_1(\Delta V_g)$ is modeled as a function proportional to the derivative of the exchange with regard to the detuning gate voltage

$$\gamma_1(\Delta V_g) = |(dJ/d\Delta V_g)/\hbar|\,\delta(\Delta V_g) \quad (6)$$

where $\delta(\Delta V_g)$ is the magnitude of the fluctuation of $\Delta V_g$, which depends on the classical control circuitry that generates the detuning gate voltage pulse. The value is typically smaller for a lower frequency pulse. Here, for simplicity, we assume $\delta(\Delta V_g) = 1\ \mu V$. The derivative parameter, $(dJ/d\Delta V_g)$, is numerically computed, as discussed in Fig. 4.

Figure 5(a) shows that increase of the detuning gate voltage $\Delta V_g$ in the range of $\Delta V_g < \Delta V_{gR}$ improves the switching speed but results in a nonmonotonically varying fidelity, due to two competing mechanisms. When the detuning gate voltage increases initially, increase of the switching speed results in an improved fidelity. However, as the detuning gate voltage further increases and approaches the resonant bias condition, the derivative $(dJ/d\Delta V_g)$ increases rapidly. The increase of the dephasing rate $\gamma_1(\Delta V_g)$ dominates over the improvement of the switching speed, and the fidelity decreases.

Scaling down the DQD spacing to $L_s = 25$ nm provides significant speed improvement by a factor of ~100×, as shown in Fig. 5(b). A switching frequency >1GHz can be achieved with a fidelity of > 90% at a detuning gate voltage of $\Delta V_g \approx$ 11.2 mV with $\frac{U(\Delta V_g)}{h_c} = 20$. Because the switching frequency is orders of magnitude higher compared to the decoherence rate, the fidelity is not limited by the decoherence mechanisms modeled in the dephasing term in the Lindblad Master equation. Instead, as the gate voltage increases and the bias approaches the resonant condition, the hybridization between the $S_{11}$ and $S_{20}$ states increases, which leads to faster speed but an increased information leakage to the $S_{20}$ state, which lowers the fidelity. Adiabatic or superadiabatic operations can not only reduce the leakage to higher state, but also lower the switching speed [20], [38]. Nevertheless, scaling down the DQD spacing to ~25 nm still provides an attractive path to achieve a fast and reliable silicon quantum gate with a low detuning gate voltage. The results in both Figs. 5(a) and (b) indicate a careful choice of the detuning bias voltage $\Delta V_g$ is essential to balance the speed and quantum fidelity.

For the modeled quantum gates in Fig. 5, the magnetic fields are chosen to make the difference of the Zeeman splitting energies between two qubits comparable to the exchange coupling energy $J$. It makes each qubit individually accessible, and at the same time, the single-qubit rotational gates sufficiently fast. Because the exchange coupling energy increases by about two orders of magnitude when $L_s$ reduces from 40 nm to 25 nm, the magnetic fields simulated also increase by two orders of magnitude. While the magnetic field gradient is large for the modeled gate with $L_s = 25$ nm, it is within the experimentally achievable range by using nanoparticles [39]

Practical applications of quantum computing need to integrate at least over 50 qubits and many quantum gates [9]. Variations of the device size parameters such as $L_s$ are inevitable. The effect is more important when the device size scales down, because $\delta L_s/L_{s0}$ can increase, where $\delta L_s$ is the standard deviation of the $L_s$ distribution, and $L_{s0}$ is the nominally designed value. To examine the variability effect, we perform a sensitivity test by a perturbation of $\Delta L_s = \pm 1$ nm, with all other parameters unchanged. Fig. 6(a) shows the quantum tomography [40] for $\Delta L_s = -1$ nm, which indicates a large deviation from an ideal CZ gate. The fidelity is reduced from 97.4% to 72.7% when $L_s$ is perturbed from 25 to 24 nm. For a variation of $\Delta L_s = 1$ nm as shown in Fig. 6(b), the fidelity reduces to 78.2%. The results indicate that a small variation of $L_s$ can lower the fidelity of the quantum gate considerably because the entanglement has an exponential dependence on the DQD spacing.

A simple quantum algorithm, the Deutsch-Jozsa algorithm, is used to further characterize the impact of the variability on the implementation of a quantum algorithm [7], [9]. The Deutsch-Jozsa algorithm determines whether a function is constant ($f_1(x) = 0$ or $f_2(x) = 1$) or balanced ($f_3(x) = x$ or $f_4(x) = 1 - x$) with a single evaluation of $U_{fi}$, which is the quantum oracle implementing the classical $f_i(x)$, as shown in Fig. 7(a). Only single-qubit quantum gates are needed for implementing the constant functions, whereas for the balanced functions $f_3(x) = x$ and $f_4(x) = 1 - x$, two-qubit quantum gates are necessary. The implementation is similar to that in [7], where a $CZ_{11}$ gate is used to implement $f_3(x) = x$, and a $CZ_{00}$ gate is used to implement $f_4(x) = 1 - x$. Here we focus on using the $CZ_{11}$ gate to implement $f_3(x) = x$. The case of $f_4(x) = 1 - x$ is similar. We assume all single-qubit quantum gates used in the implementation are ideal with 100% fidelity, so that the only source of error is the two-qubit CZ gate.

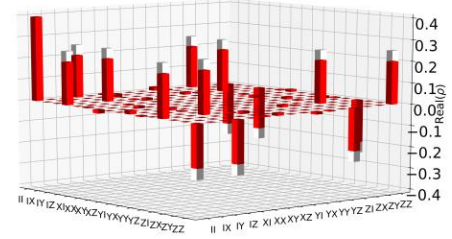

(a)

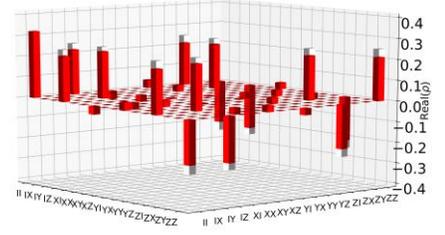

(b)

Fig. 6. Quantum process tomography of the $CZ_{11}$ gate in the presence of the DQD spacing variation for (a) $\Delta L_s = -1$ nm and (b) $\Delta L_s = 1$ nm. The white bar shows the difference to an ideal $CZ_{11}$ gate. The modeled device structure is shown in Fig. 1(c) and the nominally designed value of $L_s$ without variation is $L_s = 25$ nm.

Fig. 7(b) shows the quantum tomography of the output density matrix for a nominally designed quantum gate with $L_{s0} = 25$ nm. The first diagonal entry of the output density matrix $\rho_{00}$ is used as the error probability, because it indicates the probability that the balanced function is erroneously identified as a constant function. For the nominally designed $CZ_{11}$ gate with $L_{s0} = 25$ nm as shown in Fig. 7(b), $\rho_{00} < 5 \times 10^{-3}$ indicates a low error of <0.5%. The error, however, increases dramatically to 19.5% for the $CZ_{11}$ gate with a perturbation of $\Delta L_s = -1$ nm, as shown in Fig. 7(c). With $\Delta L_s = 1$ nm, Fig. 7(d) shows an error of 11.0% for $f_3(x) = x$. The analysis indicates that a small variation of $|\Delta L_s| = 1$ nm can create an error above 10%. For a larger quantum system that integrates many quantum gates, the distribution of $L_s$ could fail an algorithm. A possible scheme to mitigate the variability effect of the $L_s$ distribution is to adaptively design the gate pulse for each gate, where the pulsewidth is designed to compensate the deviation of $L_s$ from its nominally designed value. By using an adaptively adjusted gate pulsewidth for the cases of $\Delta L_s = 1$ nm and $\Delta L_s = -1$ nm, the simulation results indicate that the error rate can be reduced to <1.5% for implementation the Deutsch-Jozsa algorithm. The method, however, requires co-design of the quantum system with the classical control circuitry and adds complexity to the control circuitry.

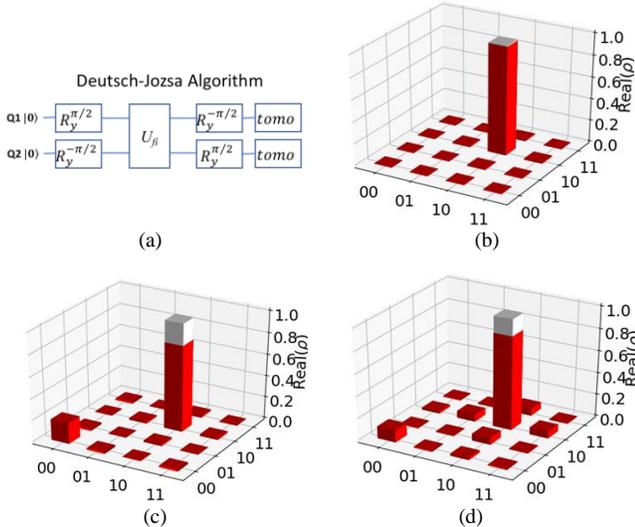

Fig. 7. Effect of the $L_s$ variation of the CZ gate on the fidelity of the Deutsch-Jozsa algorithm. (a) Diagram of implementing the Deutsch-Jozsa algorithm. The real part of the output density matrix when the balanced function $f_3(x) = x$ is implemented by a $CZ_{11}$ quantum gate designed (b) with the nominally designed $L_{s0} = 25$ nm without variation, (c) with $\Delta L_s = -1$ nm variation, and (d) with $\Delta L_s = 1$ nm variation. The white bars show the difference to the ideal case. The modeled device is in Fig. 1(c), and the applied detuning gate voltage is $\Delta V_g = 11.57$ mV with a pulsewidth of $t_{pulse} = 0.373$ ns.

IV. CONCLUSION

By developing and using a multiscale device simulation method, the performance potential and variability issues of the silicon quantum gate based on the detuning mechanism is examined. It is projected that a fast switching speed greater than 1GHz with a high fidelity greater than 90% can be achieved in a silicon-based controlled phase gate with a DQD spacing of ~25 nm, with a low operation voltage of $\Delta V_g$ ~10 mV. It is shown that the gate-voltage-induced detuning mechanism is efficient to control the entanglement in silicon quantum gate devices. At the same time, the careful choice of the bias condition is essential to balance the speed and fidelity. Scaling down the DQD spacing provides significant device performance boost in terms of the speed and fidelity. It is noted that device-to-device variability must be addressed to scale up the quantum gates for silicon-based quantum computing. A method to mitigate the variability effect by co-designing the classical control circuitry to compensate for the effect of the quantum gate size variability is proposed.

REFERENCES

[1] B. M. Maune *et al.*, "Coherent singlet-triplet oscillations in a silicon-based double quantum dot," *Nature,* vol. 481, no. 7381, pp. 344-347, Jan. 2012, doi: 10.1038/nature10707.
[2] D. Loss and D. P. DiVincenzo, "Quantum computation with quantum dots," *Phys. Rev. A*, *Gen. Phys.*, vol. 57, no. 1, pp. 120-126, Jan. 1998, doi: 10.1103/PhysRevA.57.120.
[3] D. D. Awschalom, L. C. Bassett, A. S. Dzurak, E. L. Hu and J. R. Petta, "Quantum spintronics: Engineering and manipulating atom-like spins in semiconductors," *Science,* vol. 339, no. 6124, pp. 1174-1179, Mar. 2013, doi: 10.1126/science.1231364.
[4] M. Veldhorst *et al.*, "A two-qubit logic gate in silicon," *Nature,* vol. 526, no. 7573, pp. 410-414, Oct. 2015, doi: 10.1038/nature125363.
[5] R. Maurand *et al.*, "A cmos silicon spin qubit," *Nature. Commun.,* vol. 7, Nov. 2016, Art. no. 13575, doi: 10.1038/ncomms13575.
[6] D. M. Zajac *et al.*, "Resonantly driven cnot gate for electron spins," *Science,* vol. 359, no. 6374, pp. 439-442, Jan. 2018, doi: 10.1126/science.aao5965.
[7] T. F. Watson *et al.*, "A programmable two-qubit quantum processor in silicon," *Nature,* vol. 555, no. 7698, pp. 633-637, Mar. 2018, doi: 10.1038/nature25766.
[8] M. A. Eriksson *et al.*, "Spin-based quantum dot quantum computing in silicon," *Quantum Inf. Process.,* vol. 3, nos. 1-5, pp. 133-146, Oct. 2004, doi: 10.1007/0-387-27732-3_10.
[9] M. A. Nielsen and I. L. Chuang, *Quantum Computation and Quantum Information*. Cambridge, U.K.: Cambridge Univ. Press, 2000.
[10] R. P. Feynman, "Simulating physics with computers," *Int. J. Theor. Phys,* vol. 21, nos. 6-7, pp. 467-488, Jun. 1982, doi: 10.1007/BF02650179.
[11] D. Deutsch, "Quantum theory, the church–turing principle and the universal quantum computer," *Proc. Roy. Soc. London A, Math.*, *Phys. Eng. Sci.*, vol. 400, no. 1818, pp. 97-117, Jul. 1985, doi: 10.1098/rspa.1985.0070.
[12] E. T. Campbell, B. M. Terhal, and C. Vuillot, "Roads towards fault-tolerant universal quantum computation," *Nature,* vol. 549, no. 7671, pp. 172-179, Sep. 2017, doi: 10.1038/nature23460.
[13] F. H. L. Koppens *et al.*, "Driven coherent oscillations of a single electron spin in a quantum dot," *Nature,* vol. 442, no. 7104, pp. 766-771, Aug. 2006, doi: 10.1038/nature05065.
[14] K. C. Nowack, F. H. L. Koppens, Y. V. Nazarov, and L. M. K. Vandersypen, "Coherent control of a single electron spin with electric fields," *Science,* vol. 318, no. 5855, pp. 1430-1433, Nov. 2007, doi: 10.1126/science.1148092.
[15] M. D. Shulman, O. E. Dial, S. P. Harvey, H. Bluhm, V. Umansky, and A. Yacoby, "Demonstration of entanglement of electrostatically coupled singlet-triplet qubits," *Science,* vol. 336, no. 6078, pp. 202-205, Apr. 2012, doi: 10.1126/science.1217692.
[16] J. M. Elzerman *et al.*, "Few-electron quantum dot circuit with integrated charge read out," *Phys. Rev. B*, *Condens. Matter*, vol. 67, no. 16, Apr. 2003, Art. no. 161308(R), doi: 10.1103/PhysRevB.67.161308.
[17] X. Mi *et al.*, "A coherent spin–photon interface in silicon," *Nature,* vol. 555, pp. 599-603, Feb. 2018, doi: 10.1038/nature25769.
[18] X. Mi, J. V. Cady, D. M. Zajac, P. W. Deelman, and J. R. Petta, "Strong coupling of a single electron in silicon to a microwave photon," *Science,* vol. 355, no. 6321, pp. 156-158, Jan. 2017, doi: 10.1126/science.aal2469.
[19] N. Samkharadze *et al.*, "Strong spin-photon coupling in silicon," *Science,* vol. 359, no. 6380, pp. 1123-1127, Mar. 2018, doi: 10.1126/science.aar4054.
[20] J. Schliemann, D. Loss, and A. H. MacDonald, "Double-occupancy errors, adiabaticity, and entanglement of spin-qubits in quantum dots," *Phys. Rev. B*, *Condens. Matter*, vol. 63, no. 8, Feb. 2001, Art. no. 085311, doi: 10.1103/PhysRevB.63.085311.
[21] T. Meunier, V. E. Calado, and L. M. K. Vandersypen, "Efficient controlled-phase gate for single-spin qubits in quantum dots," *Phys. Rev. B*, *Condens. Matter*, vol. 83, no. 12, Mar. 2011, Art. no. 121403R, doi: 10.1103/PhysRevB.83.121403.
[22] J. Pedersen, C. Flindt, N. A. Mortensen, and A.-P. Jauho, "Failure of standard approximations of the exchange coupling in nanostructures," *Phys. Rev. B*, *Condens. Matter*, vol. 76, no. 12, Sep. 2007, Art. no. 125323, doi: 10.1103/PhysRevB.76.125323.
[23] Q. Li, Ł. Cywiński, D. Culcer, X. Hu, and S. D. Sarma, "Exchange coupling in silicon quantum dots: Theoretical considerations for quantum computation," *Phys. Rev. B*, *Condens. Matter*, vol. 81, no. 8, Feb. 2010, Art. no. 085313, doi: 10.1103/PhysRevB.81.085313.
[24] T. Wu and J. Guo, "Performance assessment of resonantly driven silicon two-qubit quantum gate," *IEEE Electron Device Lett.*, vol. 39, no. 7, pp. 1096-1099, Jul. 2018, doi: 10.1109/LED.2018.2835385.
[25] J. R. Petta *et al*., "Coherent manipulation of coupled electron spins in semiconductor quantum dots," *Science,* vol. 309, no. 5744, pp. 2180-2184, Sep. 2005, doi: 10.1126/science.1116955.

[26] G. Lindblad, "On the generators of quantum dynamical semigroups," *Commun. Math. Phys.,* vol. 48, no. 2, pp. 119-130, Jun. 1976, doi: 10.1007/BF01608499.

[27] H.-P. Breuer and F. Petruccione, *The Theory of Open Quantum Systems*. New York, NY, USA: Oxford Univ. Press, 2002.

[28] J. J. Pla *et al*., "A single-atom electron spin qubit in silicon," *Nature,* vol. 489, no. 7417, pp. 541-545, Sep. 2012, doi: 10.1038/nature11449.

[29] E. Kawakami *et al*., "Electrical control of a long-lived spin qubit in a si/sige quantum dot," *Nature Nanotechnol.*, vol. 9, no. 9, pp. 666-670, Aug. 2014, doi: 10.1038/nnano.2014.153.

[30] J. T. Muhonen *et al*., "Storing quantum information for 30 seconds in a nanoelectronic device," *Nature Nanotechnol.*, vol. 9, no. 12, pp. 986-991, Oct. 2014, doi: 10.1038/nnano.2014.211.

[31] C. Auth *et al*., "A 22nm high performance and low-power CMOS technology featuring fully-depleted tri-gate transistors, self-aligned contacts and high density MIM capacitors," in *Proc. Symp. VLSI Technol. (VLSIT)*, Honolulu, HI, USA, Jun. 2012, pp. 131-132.

[32] M. Friesen, S. Chutia, C. Tahan, and S. N. Coppersmith, "Valley splitting theory of SiGe/Si/SiGe quantum wells," *Phys. Rev. B*, *Condens. Matter*, vol. 75, no. 11, Mar. 2007, Art. no. 115318, doi: 10.1103/PhysRevB.75.115318.

[33] C. H. Yang *et al.*, "Spin-valley lifetimes in a silicon quantum dot with tunable valley splitting," *Nature Commun.*, vol. 4, Jun. 2013, Art. no. 2069, doi: 10.1038/ncomms3069.

[34] C. J. Cramer, *Essentials of Computational Chemistry: Theories and Models*, 2nd ed. London, U.K.: Wiley, 2013, pp. 191-232.

[35] E. Nielsen, R. W. Young, R. P. Muller, and M. Carroll, "Implications of simultaneous requirements for low-noise exchange gates in double quantum dots," *Phys. Rev. B*, *Condens. Matter*, vol. 82, no. 7, Aug. 2010, Art. no. 075319, doi: 10.1103/PhysRevB.82.075319.

[36] R. Venugopal, Z. Ren, S. Datta, M. Lundstrom, and D. Jovanovic, "Simulating quantum transport in nanoscale transistors: Real versus mode-space approaches," *J. Appl. Phys.,* vol. 92, no. 7, pp. 3730-3739, 2002.

[37] M. D. Reed *et al.*, "Reduced sensitivity to charge noise in semiconductor spin qubits via symmetric operation," *Phys. Rev. Lett.,* vol. 116, no. 11, Mar. 2016, Art. no. 110402, doi: 10.1103/PhysRevLett.116.110402.

[38] B. B. Zhou *et al.*, "Accelerated quantum control using superadiabatic dynamics in a solid-state lambda system," *Nature Phys.,* vol. 13, no. 4, pp. 330-334, Nov. 2016, doi: 10.1038/nphys3967.

[39] K. Ngamchuea, K. Tschulik, and R. G. Compton, "Magnetic control: Switchable ultrahigh magnetic gradients at $Fe_3O_4$ nanoparticles to enhance solution-phase mass transport," *Nano Res*., vol. 8, no. 10, pp. 3293-3306, Oct. 2015, doi: 10.1007/s12274-015-0830-y.

[40] J. L. O'Brien *et al.*, "Quantum process tomography of a controlled-NOT gate," *Phys. Rev. Lett.,* vol. 93, no. 8, Aug. 2004, Art. no. 080502, doi: 10.1103/PhysRevLett.93.080502.